\documentclass[aps,showpacs,superscriptaddress,prb,amsmath,twocolumn,floatfix]{
revtex4}
\usepackage{graphicx}
\newcommand{\br}{\mathbf{r}}
\begin{document}
\title{Van der Waals Coefficients of Atoms and Molecules from a Simple
Approximation for the Polarizability}
\author{Huy-Viet Nguyen}	
\affiliation{SISSA-Scuola Internazionale Superiore di Studi Avanzati, 
                      via Beirut 2-4, I-34014 Trieste, Italy}
\affiliation{Physics Faculty, Hanoi National University of Education, 
                    136 Xuan-Thuy, Cau-Giay, Hanoi, Vietnam}
\author{Stefano de Gironcoli}	
\affiliation{SISSA-Scuola Internazionale Superiore di Studi Avanzati,
                     via Beirut 2-4, I-34014 Trieste, Italy}
\affiliation{INFM DEMOCRITOS National Simulation Centre, 
                     via Beirut 2-4, I-34014 Trieste, Italy}
\date{\today}
\begin{abstract}
A simple and computationally efficient scheme to calculate approximate imaginary-frequency dependent polarizability, 
hence asymptotic van der Waals coefficient, within density functional theory is proposed. 
The dynamical dipolar polarizabilities of atoms and molecules are
calculated starting from the Thomas-Fermi-von Weizs\"acker (TFvW) approximation
for the independent-electron kinetic energy functional.
The van der Waals coefficients for a number of closed-shell ions and a few molecules are hence
calculated and compared with available values obtained by fully first-principles calculations. 
The success in these test cases shows the potential of the proposed TFvW approximate response function in capturing the essence of 
long range correlations and may give useful information for constructing a functional which naturally
includes van der Waals interactions.
\end{abstract}
%
%
\maketitle
\section{Introduction}
Density functional theory (DFT)\cite{HK} has become a standard tool for electronic structure calculations
of atoms, molecules and materials thanks to the widely used local density approximation (LDA) \cite{KS} and 
generalized gradient ones (GGA) \cite{gga} to the exchange-correlation functionals
in the Kohn-Sham (KS) formulation. 
These approximations describe well many properties--such as cohesion, bonds, structures, vibrations, etc.--
of densely packed molecules and materials when covalent, metallic or hydrogen bonds are involved.
However, they fail when applied to systems having regions of small overlapping density where long range
correlation effects, which are not treated correctly in LDA and GGA, are important. 
These kinds of systems are frequently met in nature, e.g. in biomolecules, as well as in
physical and chemical processes, 
such as, for instance, in molecular adsorption on surfaces, chemical reactions, etc. 
The simplest paradigmatic examples of these weakly bonded systems are
dimers of noble gas atoms and/or closed-shell molecules and several works in the
literature exists where the van der Waals (vdW) $C_6$ coefficients are calculated within DFT at 
different levels of sophistication, ranging from a crude treatment of 
response functions in term of electronic densities \cite{ALL96} to full 
calculations within the time-dependent DFT framework. \cite{C6TDDFT} 

During the past decade, many attempts have been done to improve performances 
of DFT calculations in these systems,\cite{KohnMeirMakarov98,Ashcroft91,Dobson,vdW-DF}
by including explicitly in the exchange and correlation functional non-local correlations, 
that are missing both in LDA and in GGA.
The general staring point in all these approaches is the so-called
Adiabatic Connection Fluctuation-Dissipation (ACFD)\cite{ACFD} expression
for the exchange-correlation (xc) energy, where an exact formula is
used, giving this quantity through dynamical response functions of all
fictitious systems which connect the non-interacting KS system with
the real many-body interacting one. These calculations are however 
extremely demanding for practical applications and have been performed
only for a limited number of systems so far \cite{RPA-EC}. Recently, an
approximate scheme has been proposed where the ACFD was used as starting
point for further simplifications and which has demonstrated to be able
to account for van der Waals interaction with some success for a number
of cases.\cite{vdW-DF}

In our opinion, more calculations and further developments need to be
pursued in order to explore the potential offered by an approximate
treatment of ACFD for the calculation of accurate exchange-correlation
energies. Our aim in this work is to assess the possibility of accurately
approximate the frequency dependent response function of a 
system starting from the Thomas-Fermi and von Weizs\"acker (TFvW)
approximation for the non-interacting kinetic energy functional. 
Only the dipolar response
function responsible for the asymptotic dispersion interaction between non
overlapping fragments will be considered here. The encouraging results
obtained for the $C_6$ coefficients will give support to the possibility of
using similar procedures in the extremely demanding calculation of
accurate correlation energy based on ACFD.

The use of approximate kinetic energy functionals to evaluate dynamical
response properties of materials is not a new idea\cite{Zaremba94}
but is constantly appealing one and, for instance, Barejee and
Harbola\cite{Harbola} have recently calculated approximate vdW
coefficients for large alkali metal cluster using an TFvW approximation in
conjunction with an hydrodynamical approach to the collective excitation
of the system.

In this study we follow a different approach to the same problem, where,
starting from Thomas-Fermi and von Weizs\"acker approximation
for kinetic energy functional, we develop a simple and computationally
fast procedure to calculate imaginary-frequency-dependent
polarizability--hence the van der Waals interaction
in the asymptotic region--by Density Functional Perturbation
Theory (DFPT).\cite{Sternheimer, Mahan80,DFPT}
Our method contains the following ingredients:
(i) the ground-state (GS) electronic density, $n(\mathbf{r})$, which is accurately computed 
from the standard KS procedure within LDA or GGA;
(ii) a single auxiliary wave function, $\varphi(\mathbf{r})$, corresponding to the GS density,
and the auxiliary Hamiltonian admitting $\varphi(\mathbf{r})$ as its GS eigenfunction in the 
TFvW approximation;
(iii) the so-called modified Sternheimer equation which is used to calculate the imaginary-frequency
polarizability of the system.
This procedure can be considered as a simplification of the one proposed by Mahan \cite{Mahan82} and others
\cite{Zaremba80,Zangwill80} more than
twenty years ago for the exact--within LDA/GGA DFT--calculation of atomic and molecular polarizabilities.
The original scheme becomes very demanding for systems having a large number of electrons
while our approximate treatment, dealing only with one single auxiliary wavefunction,
can deal efficiently with systems of any number of electrons.

This scheme could be a good starting point, we believe, for developing
an approximation of the xc-energy functional which is simple enough
to allow broad applications but still capture the essence of van der
Waals energies.\cite{NHV2} Moreover, it could also be useful for calculating
vdW coefficients of fragments which are needed in some semi-empirical
treatments of vdW energy in DFT such as the damped-dispersion force
method. \cite{DDF,Murdachaew}

\section{Theory}

\subsection{Calculation of van der Waals coefficient within DFPT}
From the theoretical point of view, van der Waals coefficients are most
conveniently determined through the frequency-dependent polarizability,
$\alpha(iu)$, of the system thanks to the relation
\begin{equation}
C_6 = \frac{3}{\pi}\int_0^{\infty} du~ \alpha_1(iu)\alpha_2(iu).
\label{C6}
\end{equation}
In the framework of density functional theory, the static polarizability,
$\alpha(0)$ can be obtained from the density response of the system under a
uniform external electric field (along the z-axis) via
\begin{equation}
\alpha(0)=2\int\frac{z\Delta n(\br)}{E}d\br.
\label{alpha0}
\end{equation}
An efficient computational technique for calculation of density response
that avoids the computation of unoccupied states of the independent-particle
Hamiltonian has been proposed by Sternheimer~\cite{Sternheimer} more than fifty
years ago for atomic polarizability calculations. The method was later modified
by Mahan\cite{Mahan80} to include a self-consistent treatment of the electrons
for the calculation of atomic polarizabilities within density-functional theory in LDA. 
The linear density-response, $\Delta n$, to an external perturbation, $\Delta
V_{ext}$,
is determined by the following self-consistent set of linear equations, 
\begin{eqnarray}
&&\Delta n = 2 \sum_i {\cal{R}}e[\psi_i^{\ast}\Delta\psi_i], \\
&&\Delta V_{KS} = \Delta V_{ext} + \Delta V_{H} + \Delta v_{xc}, \\
&&\left[-\frac{\nabla^2}{2} + V_{KS} - \varepsilon_i \right] \Delta\psi_i
=- [\Delta V_{KS} - \Delta\varepsilon_i]\psi_i,
\end{eqnarray}
where the sums run on the set of occupied orbitals only.

A generalization of the method for calculation of polarizability at
a finite imaginary frequency (see below) was also made by Mahan and
was used to calculate the van der Waals coefficients for a number of
atomic systems.\cite{Mahan82} The method has later been adapted and
successfully applied to the case of extended periodic systems where it is
better known as Density Functional Perturbation Theory.\cite{DFPT} This
approach is exact within LDA DFT, however, computationally demanding
since the computational cost grows as the third power of the system
size.

\subsection{Calculation of van der Waals coefficient using TFvW functional}

One possible way to reduce the computational cost involved in the DFPT
calculation of static and dynamical response functions is to approximate
the non interacting kinetic energy with some orbital free functional,
as in the TFvW scheme.

Let us assume that the GS density distribution, $n(\mathbf{r})$, of an
atomic or molecular system has been computed accurately within the KS
scheme, employing, for instance, LDA or GGA as xc-functional, and let us
introduce the auxiliary wave function $\varphi(\mathbf{r})$, which is
normalized to unit and simply related to the electron density as
\begin{equation}
n(\mathbf{r})= N|\varphi(\mathbf{r})|^2,
\end{equation}
where $N$ is the number of electrons in the system.
The crucial point in our scheme is the assumption that the response of the system around its GS density can be 
approximated by using the TFvW functional for the kinetic energy. 
In this approximation the total energy functional in term of $\varphi(\mathbf{r})$ reads
\begin{eqnarray}
E[\varphi] &=& \alpha \int [N\varphi]^{5/3}d\mathbf{r} +
               \frac{N}{2}\int|\nabla\varphi|^2d\mathbf{r} + 
               E_H[\varphi]  + \notag \\
                &~&E_{xc}[\varphi] +E_{ext}[\varphi] -N\mu\left[\int |\varphi|^2d\mathbf{r} - 1\right],
\label{Ephi}
\end{eqnarray}
where $\alpha =  \frac{3}{10}(3\pi^2)^{2/3} $ and $\mu$ is the Lagrange
multiplier used to enforce normalization (Rydberg atomic units are used
throughout the text).  The choice of this approximation is inspired by
the fact that the vW correction term gives the exact kinetic energy in
regions where only one wave function is relevant, typically the asymptotic
region of atoms or molecules.  Moreover, the dominant contributions to
the polarizability come from the loosely bound electrons in this region,
that are thus expected to be captured in this approximation.

The corresponding Euler equation of Eq. (\ref{Ephi}) determines $\varphi(\mathbf{r})$
\begin{equation}
\left[ -\frac{\nabla^2}{2} +  V_{ext}+V_H+ v_{xc}+
\frac{k_F^2}{2} - \mu \right]\varphi(\br)=0,
\label{TFvW-Eq}
\end{equation}
where $k_F(\br)=(3\pi^2n(\br))^{\frac{1}{3}}$ is the local Fermi wave-vector.
If $V_{ext}(\br)$ is given, an approximate GS density in TFvW approximation can be obtained by solving this equation.
Here we invert the reasoning and the density, hence $\varphi(\mathbf{r})$, is assumed to be given and 
an auxiliary effective 
potential, denoted by $V_{eff}(\br)$, is constructed such that the corresponding Hamiltonian admits $\varphi(\br)$ 
as its GS eigenfunction. It can be formally written as
\begin{equation}
V_{eff}(\br) -\mu = \frac{1}{2}\frac{\nabla^2\varphi(\br)}{\varphi(\br)},
\label{veff}
\end{equation}
and $V_{eff}(\br)$ can be found from this equation once $\varphi(\br)$ is known, even if some care must be
taken in the asymptotic region where the density, and hence $\varphi(\br)$, vanishes exponentially.

The linear density-response to an external perturbation is determined
by the following self-consistent set of equations, 
\begin{eqnarray}
&&\Delta n = 2N{\cal{R}}e[\varphi^{\ast}\Delta\varphi], \\
&&\Delta V_{eff} = \Delta V_{ext} + \Delta V_{H} + \Delta v_{xc} +
                      \frac{k_F^2}{3n}\Delta n, \\
&&\left[-\frac{\nabla^2}{2} + V_{eff} - \mu \right] \Delta\varphi =
- [\Delta V_{eff} - \Delta\mu]\varphi,
\end{eqnarray}
where
$\Delta\mu = \langle \varphi|\Delta V_{eff}|\varphi\rangle$, and  $\Delta\varphi$,
which satisfies the orthogonal condition $\langle \varphi|\Delta \varphi\rangle = 0$, 
is the first order change of the wave function brought about by the external field. 
Once this set of equations is solved, static polarizability is calculated from
Eq.~(\ref{alpha0}). In order to find the polarizability at finite imaginary
frequency, $\alpha(iu)$, the above procedure is slightly modified by adding a
frequency term $iu$ to $\mu$, making it become a complex quantity $\mu + iu $.
\cite{Mahan82}
From the imaginary-frequency polarizability, $\alpha(iu)$, the vdW coefficient $C_6$ can be immediately calculated from Eq.~(\ref{C6}).
\subsection{Construction of the effective potential}
We have chosen to apply our method to calculate dynamic polarizabilities and
vdW coefficients for two kinds of systems, namely spherically symmetric ions
and some simple molecules to demonstrate the efficiency of the method.
By exploiting the symmetry properties of the former systems, the GS density
$n(\br)$ can be calculated very accurately by integrating the radial KS
equations on a logarithmic grid with the highly accurate Numerov's algorithm.
In this case the effective potential $V_{eff}(\br)$ can be obtained
directly from Eq.~(\ref{veff}). For the latter, the situation is more
complicated because all quantities are calculated within the plane-wave
pseudopotential method and are expanded in Fourier components up to a given
kinetic-energy cut-off which makes the inversion needed in Eq. \ref{veff}
numerically difficult. We have overcome this difficulty by an {\em iterative
optimization process} inspired by the method described in
Ref.~\onlinecite{perdew03}, for the determination of the optimized effective
potential starting from an accurate density: assuming that at a given
iteration, $i^{th}$, the approximate effective potential is $V^i_{eff}(\br)$ a
residual quantity $S^i(\br)$ is defined by
\begin{equation}
S^i(\br) = \varphi^{\ast}(\br)\left[ -\frac{\nabla^2}{2}+V^i_{eff}(\br)-\mu \right] \varphi(\br) .
\end{equation}
This quantity vanishes everywhere only if the Hamiltonian corresponding to this potential
admits $\varphi(\br)$ as its GS eigenfunction. As long as this is not the case the
potential is updated as
\begin{equation}
V^{i+1}_{eff}(\br) = V^{i}_{eff}(\br) + \alpha S^i(\br) + \beta,
\end{equation}
where $\alpha$ and $\beta$ are chosen in such a way that the norm of the new residual
$ ||S^{i+1}|| = \int d\br ~[S^{i+1}(\br)]^2 $ is minimized. 
The process is terminated when the integrated charge difference 
$\delta n^i = \frac{1}{N}\int d\br ~|n^i(\br) - n(\br)|$
is less than a given threshold, practically chosen to be of the order of $10^{-3}$ .
\section{Results}
\subsection{Atomic systems}
\begin{figure}[t]
\centerline{\includegraphics[width=0.45\textwidth]{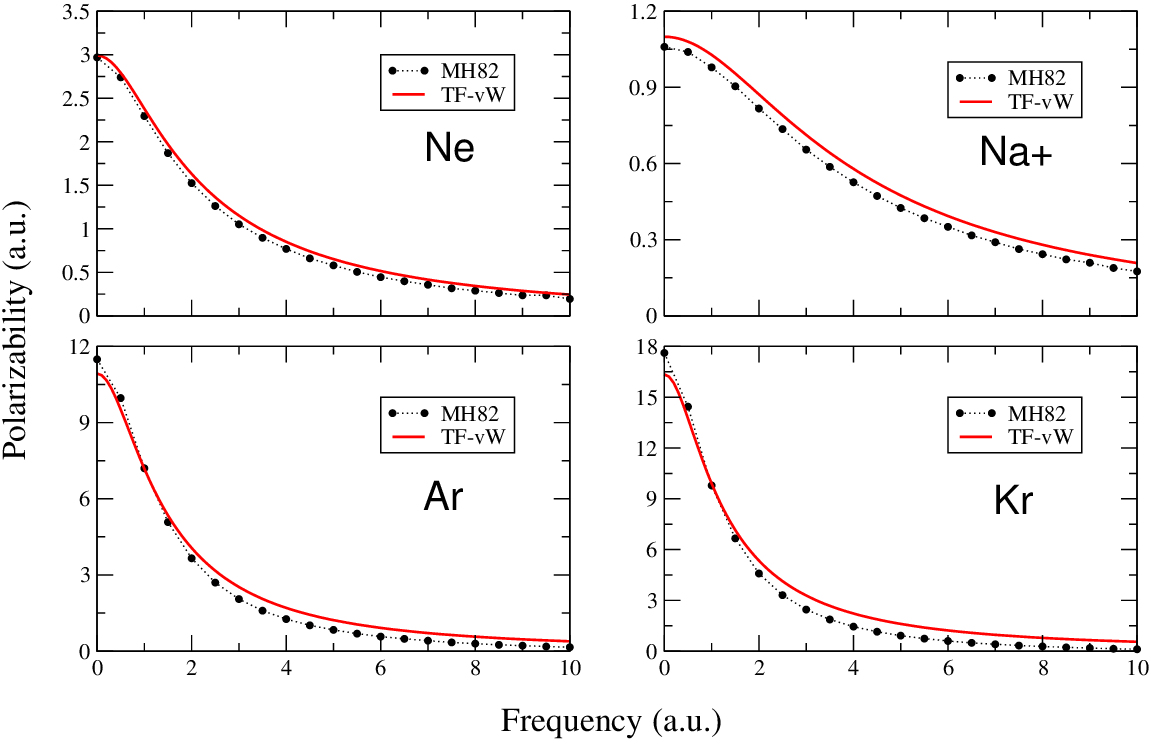}}
\caption{\label{fig:pol-compare}
Dynamic polarizabilities of several arbitrarily chosen ions calculated by our
method compared to corresponding quantities in Ref. \onlinecite{Mahan82}. }
\end{figure}
Numerical results for closed-shell ions show a good agreement with those obtained by Mahan with
an accurate but more expensive calculation for a wide range of atomic number and frequency. 
This can be seen in Fig. 1 where frequency-dependent polarizabilities of some closed-shell 
ions obtained by the two methods are compared. In Table \ref{c6} calculated values of vdW coefficients 
for a number of pairs of rare gases are given. Our results are all in the range and at least as accurate 
as those reported in Ref. \onlinecite{unified} when compared to the reference ones. 

In Fig. 2 our calculated $C_6$ values both for homonuclear and mixed pairs of $14$ ions are plotted
against those reported in Ref. \onlinecite{Mahan82}. The good agreement between the results obtained by the two methods
are indicated by the narrow
spread of the points around the diagonal. Quantitatively, the difference
never exceeds $25\%$ and in most cases is less than $10\%$.

\begin{table}
\caption{\label{c6}
$C_6$ values  for  dimers (Rydberg atomic units). Present: our results.
Mahan: calculations from  a very similar scheme in Ref. \onlinecite{Mahan82}. 
Unified: calculations from Ref. \onlinecite{unified} using 
self-consistent electrodynamics. Reference: values cited as reference in
Ref. \onlinecite{unified}. }
\begin{tabular}{lccccc }
\hline\hline
Dimer~~~         & Present~~~                   &Mahan~~~          & Unified~~~ 
                  & Reference\\
\hline
He~~~	   	&2.64~~~		&3.64~~~	&2.58~~~	
&2.92~~~\\
Ne~~~	        &15.44~~~	        &13.96~~~	&15.0~~~	
&13.8~~~	\\
Ar~~~		&133~~~		&132.2~~~	&143~~~		&134~~~	\\
Kr~~~		&266~~~		&261.4~~~	&291~~~		&266~~~	\\
Xe~~~		&600~~~		&---~~~   	&663~~~		&597~~~	\\
\hline
\end{tabular}
\end{table}

\begin{figure}[h!]
.\vskip 1.0truecm
\centerline{\includegraphics[width=0.45\textwidth]{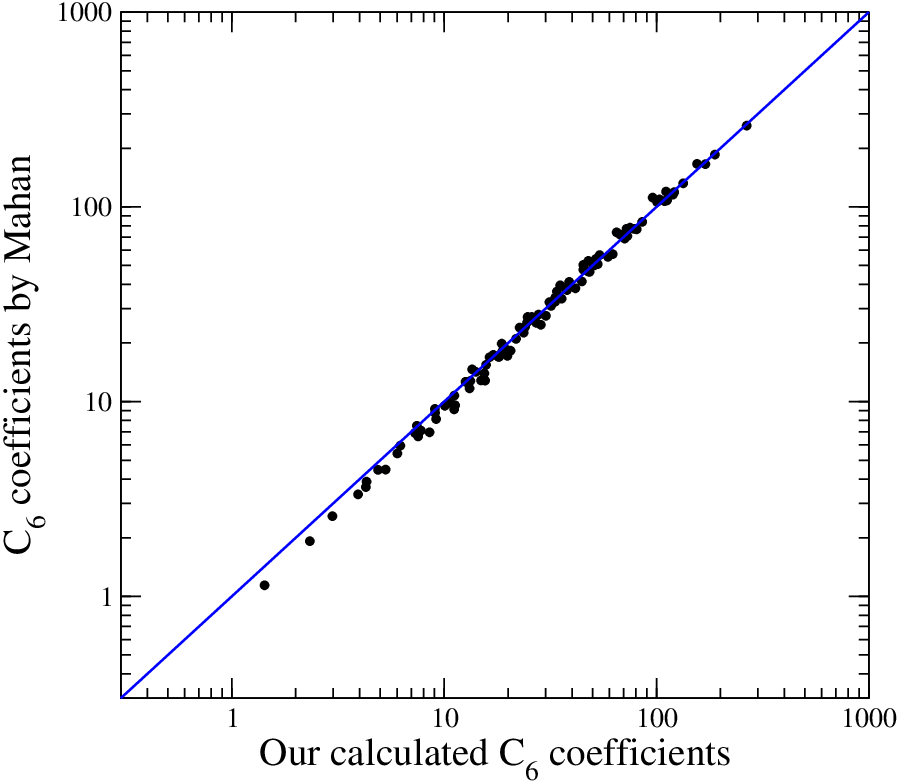}}
\caption{\label{fig:c6-compare}
$C_6$ values of all possible pairs of $14$ ions  calculated by our method 
plotted against corresponding values shown in Ref. \onlinecite{Mahan82}}
\end{figure}
\subsection{Importance of using a good charge density}
The calculation of van der Waals coefficients presented above is from
response functions in TFvW approximation calculated around an accurate
charge density. For this purpose, we need to introduce an effective
potential admitting the square root of the charge density as its ground
state wave function as already discussed in Sec. II B. One may wonder
if this is indeed necessary since it seems natural to calculate the
charge density also in TFvW approximation and then use it as input for
the calculation of $C_6$ coefficient. Moreover, doing calculation in
this way makes the construction of the effective potential needed in our
calculation unnecessary because it is determined in the self-consistent
solution of Eq. (\ref{TFvW-Eq}). We have tried this option for the case
of noble gas atoms and the results is disastrous. For example, $C_6$
coefficient of He changes from  2.1 a.u. when computed with accurate
LDA-DFT charge density to 15.1, and 227.2 (a.u.) when calculated from
the charge densities obtained from the solution of the Hartree equation
or from solving self-consistently the TFvW approximation. Reducing
the weight of the gradient corrected term in the functional to $1/5$, an
empirical value often used in the literature\cite{Stich82,Gross-Dreizler},
still gives a very poor result (36.4 a.u.).  This behaviour is not totally
unexpected since it is well-known\cite{Gross-Dreizler} that also the TFvW
kinetic energy itself -- the quantity on which the approximate response
function is based -- while giving accurate estimates when applied to
accurate charge density behaves poorly if treated self-consistently.
This result indicates the importance of calculating the response functions
with accurate charge densities and shows that our approach, though not
being a self-consistent procedure, is the correct way to calculate vdW
coefficient using TFvW approximation.
\subsection{Role of core electrons}
\begin{figure}[t]
\centerline{\includegraphics[width=0.45\textwidth]{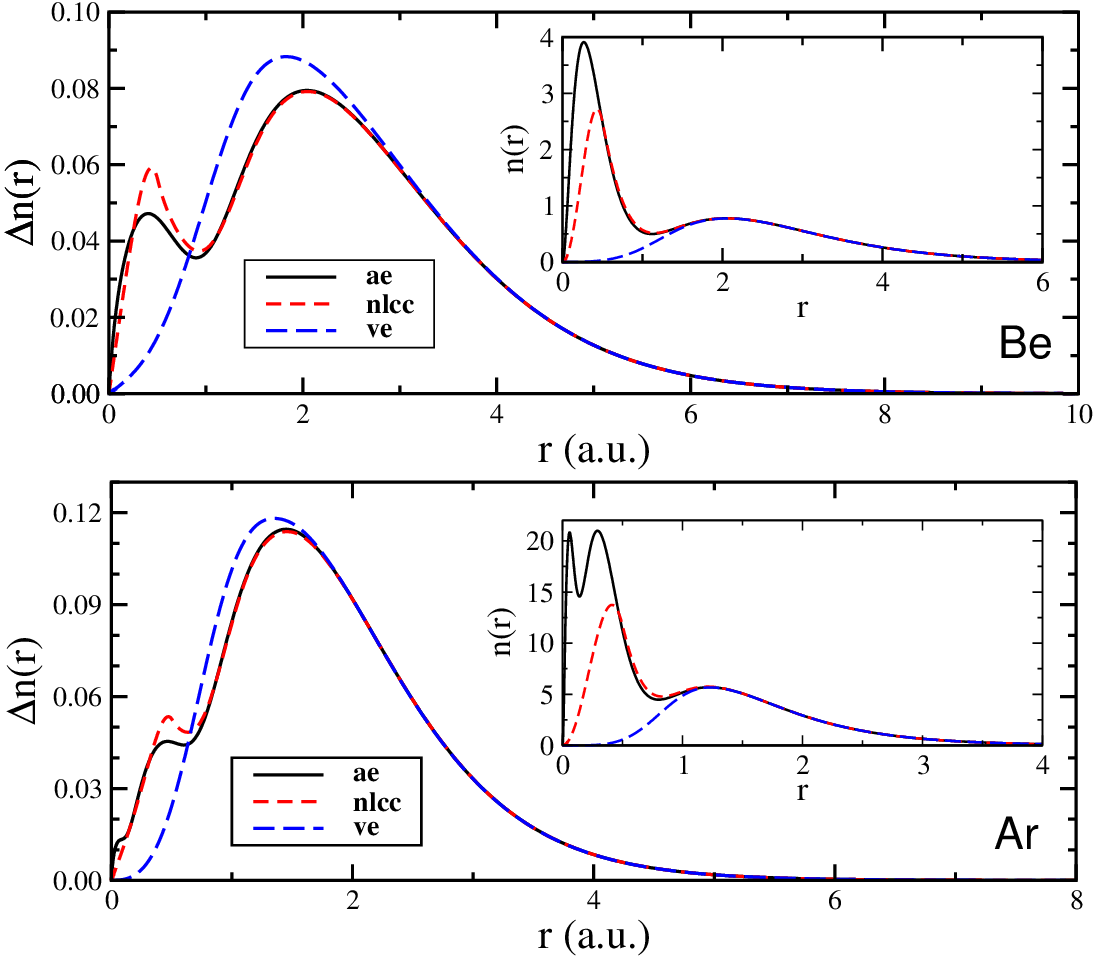}}
\caption{\label{fig:core-effect}
Static density responses of Be (top) and Ar (bottom) calculated
with different treatment of the core charge (see the insets): totally
included (ae), completely neglected (ve), and partially included (nlcc).}
\end{figure} 
For systems without spherical symmetry, the KS equations
as well as the modified Sternheimer ones are no longer radial-angular
separable, and a general method to calculate GS electronic structures must
be used to solve them. The scheme described above have been implemented
in the {\tt PWscf} plane-wave pseudopotential code which is part of
the {\em Quantum ESPRESSO} distribution \cite{espresso}. Although
in the pseudopotential approach the core-electron density could
be easily included in the definition of the auxiliary function,
$\varphi(\mathbf{r})$, it is not convenient because of the higher
computational cost due to larger kinetic-energy cutoff that would be
required in the calculation of the GS density $n(\br)$.  It is therefore
worthwhile to estimate the contribution, expected to be small, of the core
electrons to the polarizability in this scheme. To this end, we compared
the results obtained in atomic calculations where $\varphi(\mathbf{r})$
was computed from the total charge density or from the valence-only
charge density or, as it is done in the non-linear core correction
(nlcc), from the valence charge density plus a smoothed core charge.
Here we show the results for Beryllium and Argon atoms for which the
effects of the core charge on the total polarization are expected to
be considerably different. It could be not negligible for the former
due to non-tightly bound core electrons to a small charge nucleus, while it
should be very small for the later whose core electrons are more tightly
bound. It turns out that in both cases the contribution of the core
charges to $C_6$ values is very small.  In Fig. \ref{fig:core-effect}
static response densities calculated by using different densities in
the core region (see the insets) are plotted.  This figure shows two
opposite roles of core electrons in determining the polarizability.
On the one hand, they contribute to the density response in the core
region, thus making the total polarizability larger.  They prevent, on the
other hand, penetration of valence electrons into the core region when an
electric field is applied, therefore reducing the total polarizability.
\begin{table}[b]
\caption{ \label{core-effect}
Static polarizabilities $\alpha(0)$ and vdW coefficients $C_6$ (Ry atomic units)
of Be and Ar calculated with different densities: all-electron (ae), valence 
with some core-charge (nlcc), and only valence (ve) density.}
\begin{tabular}{l c c c c c c c c}
\hline\hline
                     &          &                 & $\alpha(0)$ &        &      
    &                  &$C_6$            &\\
\hline
                     &~~~~ & ae             &nlcc          &ve       &~~~~  & ae
             &nlcc               &ve \\
\hline
Be                  &~~~~ & 33.07         &33.05        &33.54  &~~~~  & 194.10 
       &194.14           &190.58 \\
Ar                   &~~~~ & 10.93         &10.92        &10.96  &~~~~  & 66.80 
         &66.85            &63.80 \\
\hline 
\end{tabular}
\end{table}
Although these contributions do not cancel each other exactly, they still
make the effects of the core charge not very important as indicated by
the values of static polarizabilities and $C_6$ coefficients reported in
Table \ref{core-effect}, where the difference between totally included
and completely neglected core charge is just a few percent.  Therefore,
in the calculations for molecules presented below, a little accuracy
has been sacrificed by using only the density of valence electrons in
order to reduce the number of plane-wave needed to describe the auxiliary
wavefunction of the system.

\subsection{Molecular systems}
To exemplify the general scheme, we have calculated dynamic
polarizabilities and vdW coefficients of methane and benzene, two
molecules with different nature of chemical bonds and geometric
structures. The KS equations for each isolated molecule  were solved
using periodic boundary conditions in a simple cubic simulation cell
with side length of $12$ and $10$ \AA~and kinetic-energy cutoffs of $80$
and $60$ Ry, respectively. Simple LDA exchange-correlation functional
with norm-conserving pseudopotentials were used to obtain the GS charge
density of the isolated molecules.

\begin{figure}[t!]
\includegraphics[width=0.45\textwidth]{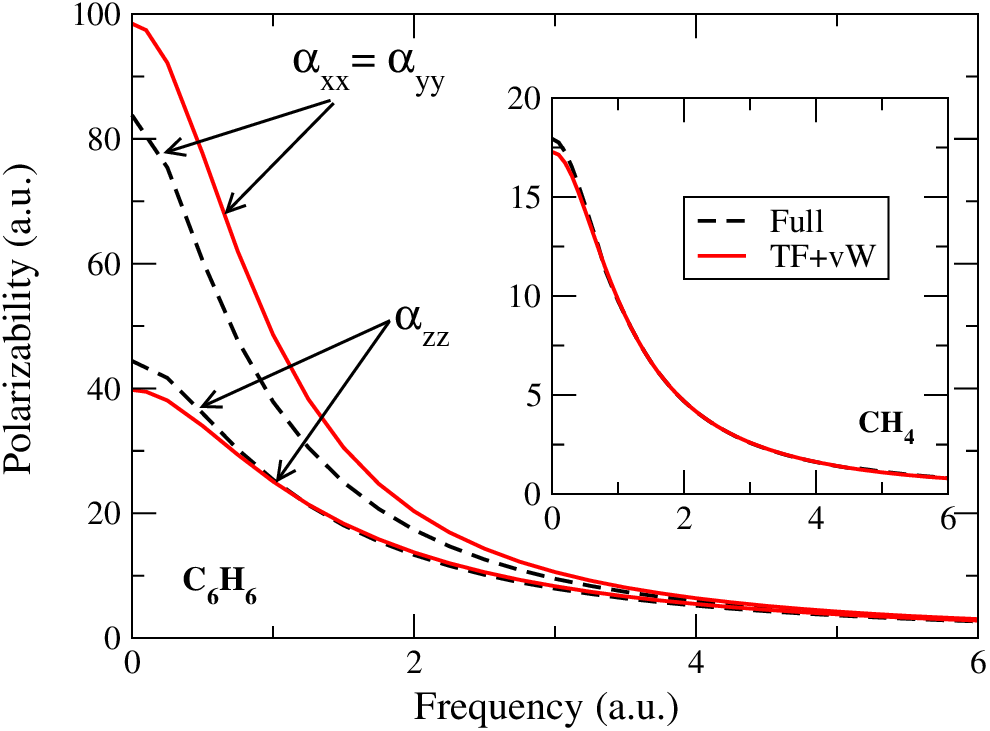}
\caption{\label{fig:me-ben}
Imaginary-frequency dependent polarizabilities of methane (the inset) and
benzene molecules calculated by TFvW method (solid red curves) compared to
results of full calculation (dashed black curves).}
\end{figure}

Fig. \ref{fig:me-ben} shows the imaginary-frequency dynamic
polarizabilities of methane and benzene molecules calculated in our
scheme, compared with the result of the full calculation which has also
been implemented in the {\tt PWscf} plane-wave pseudopotential code.
For methane molecule, the result of our simplified calculation compares
excellently with the one of the more accurate method. Although this is
not the case for benzene molecule, nevertheless the difference between
the two calculations is still rather moderate as expected.

From the dynamical polarizability the $C_6$ coefficients can be obtained
and the approximate calculation agrees very well with the full calculation
for methane (264 a.u. to be compared with 271 a.u. obtained in the full
calculation) while for benzene TFvW approximation overestimate by about 40\%
the full calculation (4.9 $\times 10^3$ a.u. w.r.t. 3.6 $\times 10^3$ a.u.).
\section{Discussion and Conclusion}

In this study we have shown that a simple approximation for the
(dynamical) polarizability of atoms and molecules can be obtained starting
from the TFvW approximation to the independent electron kinetic energy
functional. The success of the present approximate method in the calculation
of frequency dependent polarizability and vdW coefficients of atomic
and molecular systems, together with its computational efficiency,
especially for large systems, suggests that it may be a useful tool to
explore the behavior of systems where non-local long-range correlations
are important.

An algorithmic development of time-dependent density
functional theory in real time has recently been proposed by Marques and
co-workers\cite{Marques} that allows for an efficient evaluation of van
der Waals coefficients from the full Kohn-Sham response function. The
computational cost for the calculation of C$_6$ coefficients in this
approach grows quadratically with the system size. In our approximate
scheme instead the computational cost grows only linearly with the
size of the system. This is because only one auxiliary wave function
is needed no matter how many electrons are present in the system. Even
for the small atomic and molecular systems considered in this work the
computational time required by the simplified calculation is at least one
order of magnitude lower than that of the full calculation and this is
expected to become increasingly more convenient for larger systems. Only
a systematic comparison of the results obtained for more systems within
our approximate scheme as well as within other approaches will allow 
to better assess the computational and physical merits and limitations
of the various approaches.

One of the authors (HVN) would like to thank the Abdus Salam ICTP,
where the initial part of this work was done, for financial support in
the framework of ICTP/SISSA Joint Master's Degree Programme (2003-2005)
and the hospitality at the centre.

\end{document}